\documentclass{iopart}
\usepackage{graphicx}

\begin{document}

\title[K$^0_s$K$^0_s$ correlations in 7 TeV $pp$ collisions from ALICE]{K$^0_s$K$^0_s$ correlations 
in 7 TeV $pp$ collisions from the ALICE experiment at the LHC
}

\author{T. J. Humanic for the ALICE Collaboration}

\address{Department of Physics, The Ohio State University, Columbus, Ohio, USA}
\ead{humanic@mps.ohio-state.edu}
\begin{abstract}
Identical neutral kaon pair correlations are measured in 7 TeV $pp$ collisions in the ALICE experiment. K$^0_s$K$^0_s$ correlation functions are formed in 3 multiplicity $\times$ 4 $k_T$ bins. The femtoscopic kaon source parameters $R_{inv}$ and $\lambda$ are extracted from these correlation functions by fitting a femtoscopy
$\times$ PYTHIA model to them,  PYTHIA accounting for the non-flat baseline found in $pp$ collisions. Source parameters are obtained from a fit which includes quantum statistics and final-state interactions of the $a_0/f_0$ 
resonance. K$^0_s$K$^0_s$ correlations show a systematic increase in $R_{inv}$ for increasing multiplicity bin and decreasing $R_{inv}$ for increasing $k_T$ bin as seen in $\pi\pi$ correlations in the $pp$ system, as well as seen in heavy-ion collisions. Also, K$^0_s$K$^0_s$ correlations are observed to smoothly extend this $\pi\pi$ $R_{inv}$ behavior for the $pp$ system up to about three times higher $k_T$ than the $k_T$ range measured in $\pi\pi$ correlations.

\end{abstract}


\section{Introduction}
In this proceedings we present preliminary results from a K$^0_s$K$^0_s$
femtoscopy study by the ALICE experiment \cite{Aamodt:2008zz} in 7 TeV $pp$ 
collisions from the CERN LHC. The main motivations to carry out K$^0_s$K$^0_s$ femtoscopic
studies to augment the usual identical charged $\pi\pi$ femtoscopic studies are 1) to extend
the $k_T$ range of the charged $\pi\pi$ studies, 2) since K$^0_s$ is uncharged, it is not
necessary to apply a final-state Coulomb correction to the pairs as is necessary for charged $\pi\pi$ pairs, and 3) one can in principle
obtain complementary information about the collision 
interaction region by using different types
of mesons. 
Previous K$^0_s$K$^0_s$ studies have been carried
out in LEP $e^+ e^-$ collisions \cite{Abreu:1996hu, Schael:2004qn, lep3}, HERA 
$ep$ collisions \cite{Chekanov:2007ev} and
RHIC Au-Au collisions \cite{Abelev:2006gu}. Due to statistics limitations,
a single set of femtoscopic source parameters, i.e. $R_{inv}$ and $\lambda$, was
extracted in each of these studies. To our knowledge, the present study is the first
femtoscopic K$^0_s$K$^0_s$ study to be carried out a) in $pp$ collisions and b) in more than
one multiplicity and $k_T$ bin.

\section{Experimental details and results}
\begin{figure}
\begin{minipage}[t]{6.0cm}
\includegraphics[width=63mm]{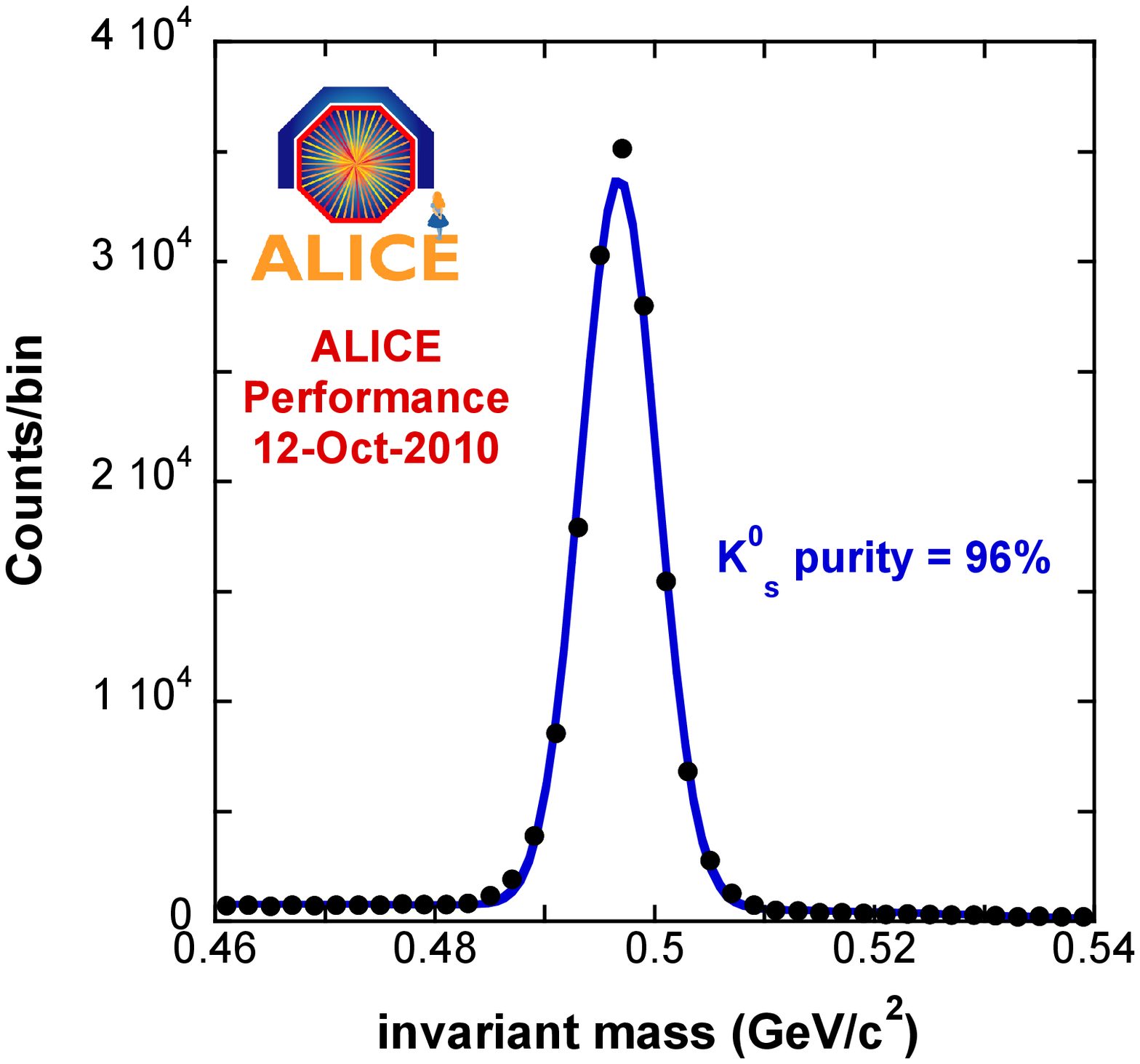} \caption{K$^0_s$ invariant mass peak.}
\label{fig1}
\end{minipage}
\hfill
\begin{minipage}[t]{6.0cm}
\includegraphics[width=60mm]{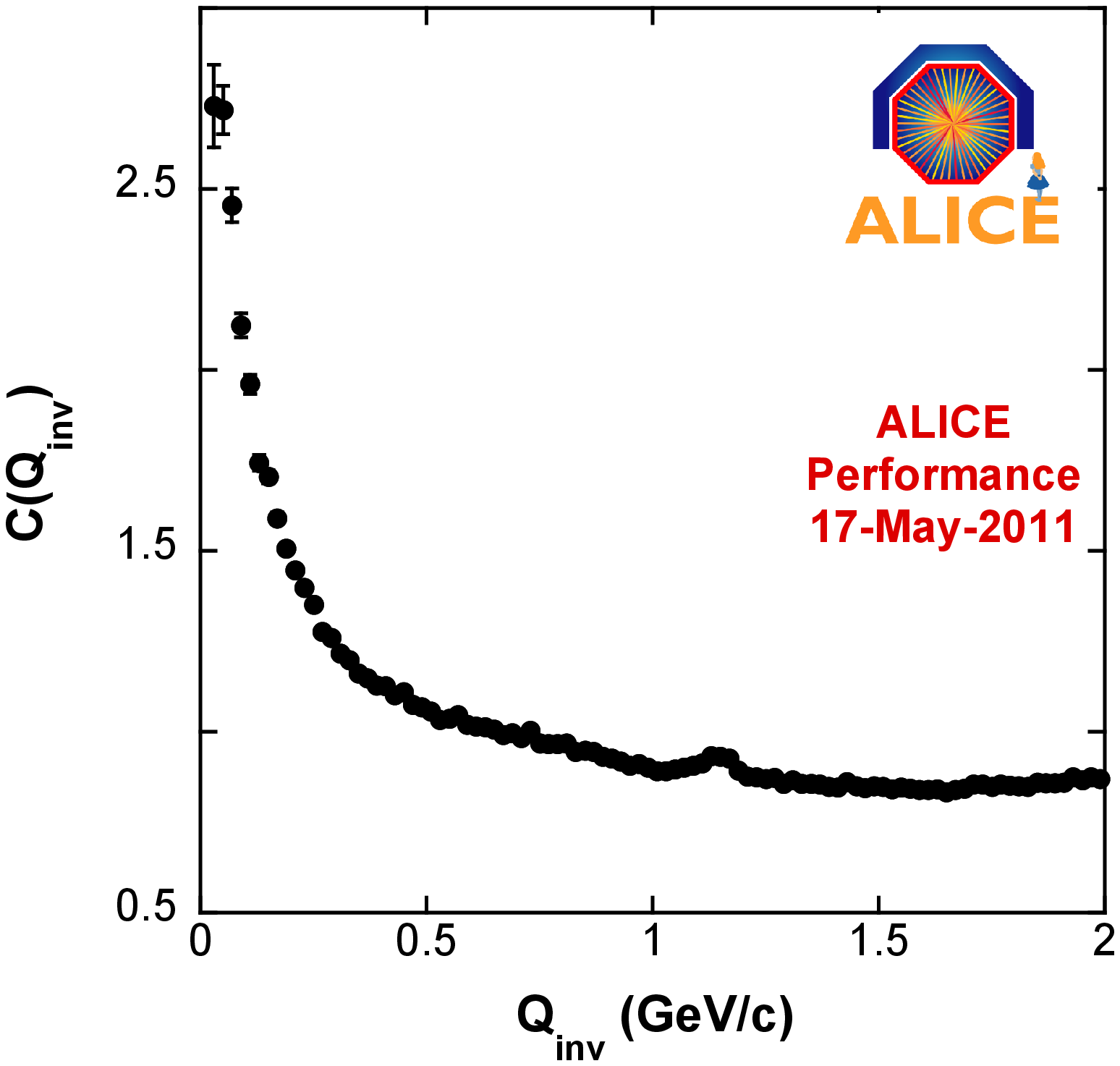} \caption{K$^0_s$K$^0_s$ $Q_{inv}$
correlation function for all multiplicity and $k_T$.}
\label{fig2}
\end{minipage}
\end{figure}

K$^0_s$ identification and momentum determination were carried out with particle
tracking in the ALICE Time Projection Chamber (TPC) and ALICE Inner 
Tracking System (ITS) \cite{Aamodt:2008zz}.
The decay channel K$^0_s\rightarrow\pi^+\pi^-$ was used for particle identification.
Figure \ref{fig1} shows an invariant mass plot of candidate K$^0_s$ vertices 
for all event multiplicities and $k_T$ along with
a Gaussian + quadratic fit to the data. A vertex was identified with a K$^0_s$ if the
invariant mass of the candidate $\pi^+\pi^-$ pair associated with it
fell in the range 0.490-0.504 GeV/$c^2$. As seen in Figure \ref{fig1},
the K$^0_s$ purity in this case is found to be 96\%. K$^0_s$ purities of similar values
are also found in the other multiplicity and $k_T$ bins used in this study.

Figure \ref{fig2} shows a K$^0_s$K$^0_s$ correlation function in the invariant
momentum difference variable 
$Q_{inv}=\sqrt{Q^2-Q_0^2}$, where $Q$ and $Q_0$ are the 3-momentum and
energy differences between the two particles respectively, 
for all multiplicity and $k_T$. The three main features
seen in this correlation function are 1) a well-defined enhancement region for
$Q_{inv}<0.3$ GeV/c, 2) a non-flat baseline for $Q_{inv}>0.3$ GeV/c 
and 3) a small peak at $Q_{inv}\approx1.15$ GeV/c. 
Considering feature 3) first, fitting a quadratic + Breit-Wigner
function to the invariant K$^0_s$K$^0_s$ mass ($m_{inv}$) distribution around this peak,
where $m_{inv}=2\sqrt{(Q_{inv}/2)^2+m_{K0s}^2}$,
we obtain
a mass of $1518 \pm 1$ MeV/$c^2$ and width ($\Gamma$) of $67\pm9$ MeV/$c^2$. Comparing
with the Particle Data Group meson table \cite{pdg}, this 
peak is a good candidate for the $f_2'(1525)$
meson. This would be the first observation of the decay of this meson into the K$^0_s$K$^0_s$
channel in $pp$ collisions. In order to disentangle the non-flat baseline from the low-$Q_{inv}$
femtoscopic enhancement, PYTHIA \cite{pythia6.4} was used to model
the baseline. Both Gaussian and quadratic fits were made to PYTHIA-generated K$^0_s$K$^0_s$
correlation functions for each multiplicity-$k_T$ bin studied, and these fits were then used
to define the baseline in fitting the femtoscopic function to the experimental correlation functions.
PYTHIA with the Perugia-0 tune \cite{perugia} is found to describe reasonably well the dependence of the baseline shape of the K$^0_s$K$^0_s$ correlation function on 
multiplicity-$k_T$ bin in the $Q_{inv}$ fitting range used of 0-1 GeV/c.
Using both the Gaussian and quadratic PYTHIA fits 
helped to define the systematic error of this
approach, as well as adding to the systematic error estimate a $\pm10$\% uncertainty
in the fit parameters due to the uncertainty in using PYTHIA to estimate the baseline.

\begin{figure}
\begin{minipage}[t]{6.0cm}
\includegraphics[width=60mm]{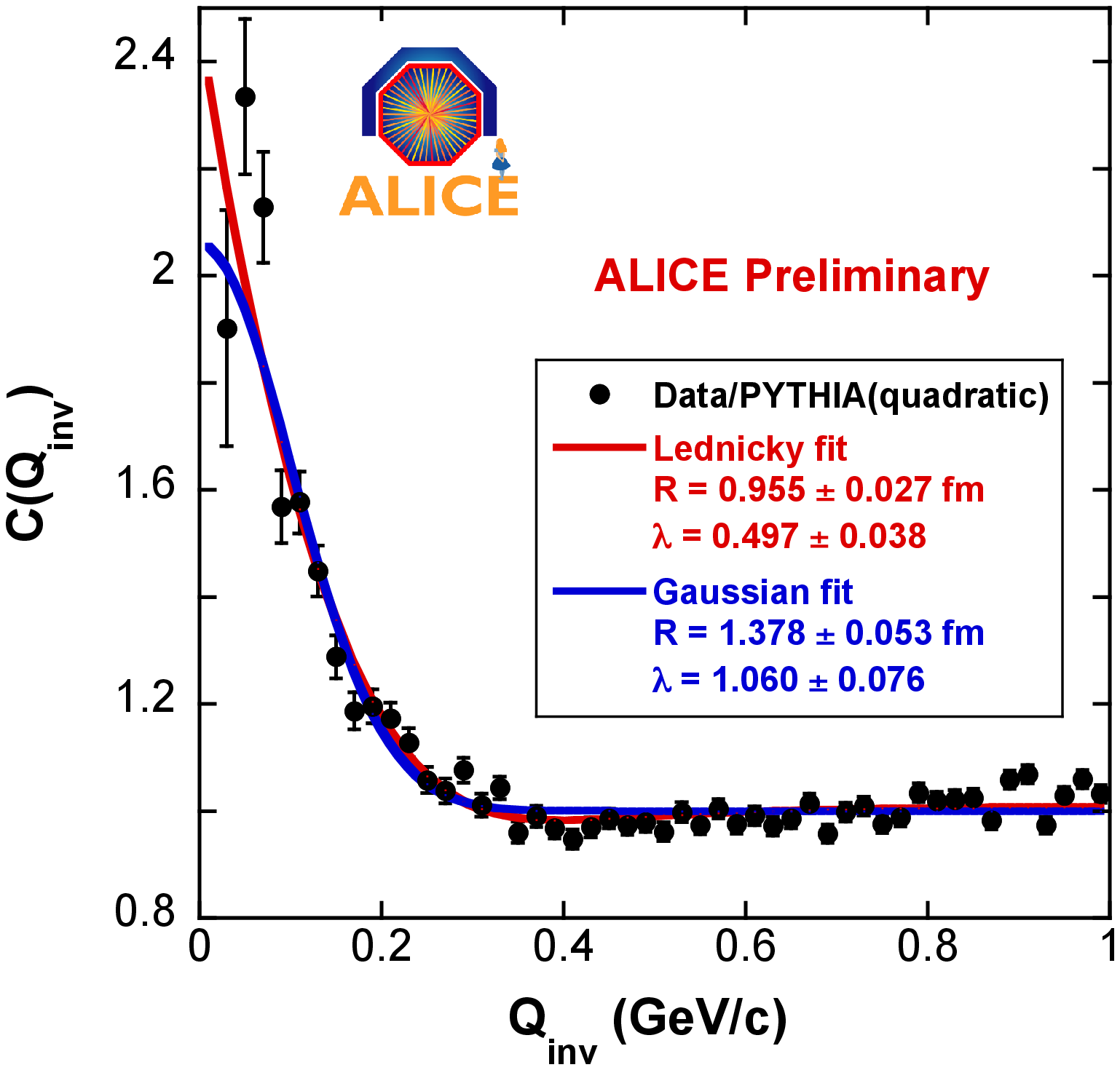} \caption{K$^0_s$K$^0_s$ correlation function
divided by the PYTHIA quadratic baseline fit for the multiplicity 1-11 and $k_T$ 0.4-0.8 GeV/c bin
with femtoscopic fits.}
\label{fig3}
\end{minipage}
\hfill
\begin{minipage}[t]{6.0cm}
\includegraphics[width=59mm]{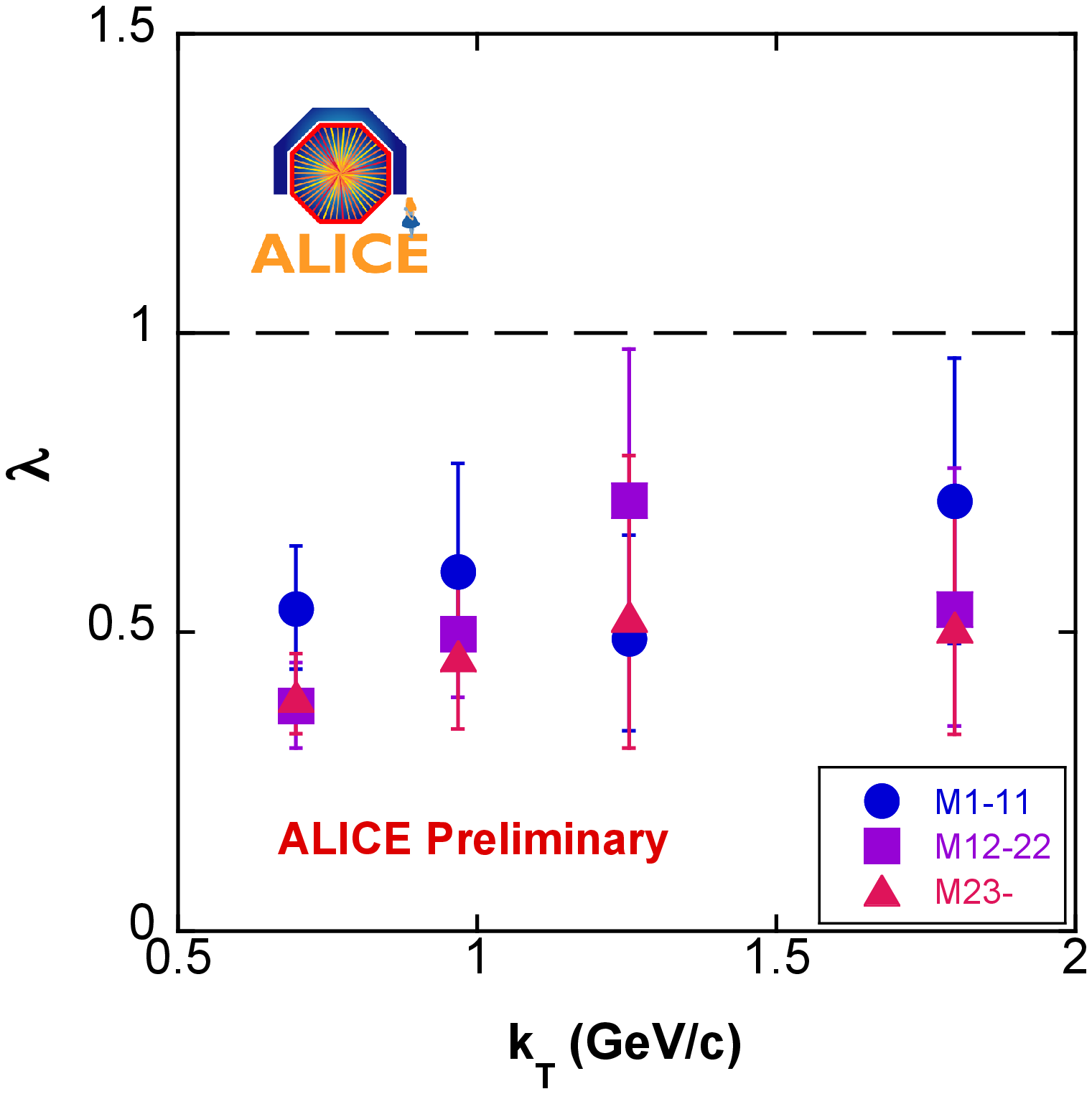} \caption{$\lambda$ parameters extracted by fitting the
Lednicky parameterization to the K$^0_s$K$^0_s$ correlation function in each of the 12 bins. Statistical + systematic
errors are shown.}
\label{fig4}
\end{minipage}
\end{figure}

K$^0_s$K$^0_s$ correlation functions in $Q_{inv}$ were formed from the data in 12 bins:
3 event multiplicity (1-11, 12-22, $\ge23$) bins $\times$ 4 $k_T$ (0.4-0.8, 0.8-1.1, 1.1-1.4,
$\ge1.4$ GeV/c) bins. The femtoscopic variables $R_{inv}$ and $\lambda$ were extracted
in each bin
by fitting the experimental correlation function divided by the PYTHIA baseline shape
by the Lednicky parameterization \cite{Abelev:2006gu} based on the
model by R. Lednicky and V.L. Lyuboshitz \cite{lednicky}.
This model takes into account both quantum statistics and strong final-state interactions
from the $a_0/f_0$ resonance which occur between the K$^0_s$K$^0_s$ pair. The K$^0_s$ 
spacial distribution is assumed to be Gaussian with a width $R_{inv}$ in the parameterization and so its
influence on the correlation function is from both the quantum statistics and the strong final-state
interaction. This is the same parameterization as was used by the RHIC STAR collaboration to
extract $R_{inv}$ and $\lambda$ from their K$^0_s$K$^0_s$ study 
of Au-Au collisions \cite{Abelev:2006gu}. Figure \ref{fig3} shows a sample experimental
K$^0_s$K$^0_s$ correlation function
divided by the PYTHIA quadratic baseline fit for the multiplicity 1-11 and $k_T$ 0.4-0.8 GeV/c bin.
Also shown are fits with the Lednicky parameterization and for comparison a usual Gaussian model. As seen,
the $R_{inv}$ and $\lambda$ parameters from the Lednicky fit are $\sim30$\% and
$\sim50$\% smaller, respectively, than those extracted from the Gaussian model fit. The
RHIC STAR experiment saw $\sim20$\% reductions in these parameters in 
Au-Au collisions compared with a Gaussian model fit \cite{Abelev:2006gu}, the smaller
effect being expected for a larger K$^0_s$ source size since the strong final-state interactions
are reduced in that case by geometry \cite{lednicky}.

\begin{figure}
\begin{center}
\includegraphics[width=128mm]{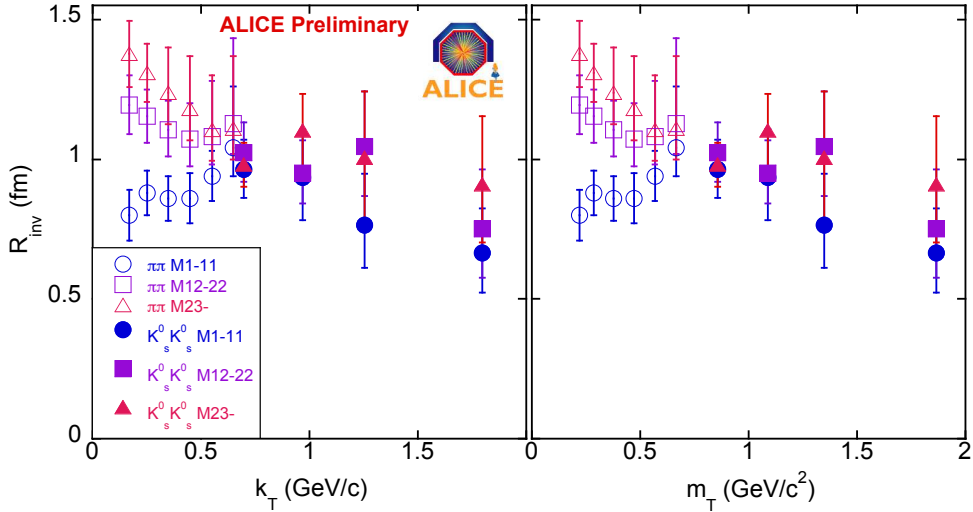} \caption{$R_{inv}$ radius parameters extracted by fitting the
Lednicky parameterization to the K$^0_s$K$^0_s$ correlation function in each of the 12 bins vs. $k_T$ and $m_T$. Also shown for comparison are $R_{inv}$ parameters extracted in the same event
multiplicity bins from a $\pi\pi$ femtoscopic study by ALICE \cite{Aamodt:2011kd} in 7 TeV $pp$ collisions. Statistical + systematic errors are shown.}
\label{fig5}
\end{center}
\end{figure}

Figures \ref{fig4} and \ref{fig5} present the results of this study for $\lambda$ and $R_{inv}$
parameters extracted by fitting the Lednicky parameterization to the K$^0_s$K$^0_s$ correlation function in each of the 12 bins. Statistical + systematic errors are shown. The largest contribution to the error bars in
all cases is due to the $\pm10$\% systematic uncertainty applied to the 
PYTHIA baseline fit parameters as
discussed above. The $k_T$ dependence of $\lambda$ is seen in Figure \ref{fig4} to be
mostly flat with $\lambda$ lying at an average level of $\sim0.5-0.6$, similar to that found in 
the ALICE $\pi\pi$ results for 7 TeV $pp$ collisions \cite{Aamodt:2011kd}. There is also a slight
tendency seen for $\lambda$ to be larger in the smaller multiplicity bins. Figure \ref{fig5} plots
$R_{inv}$ vs. both $k_T$ and $m_T$. Also shown for comparison are $R_{inv}$ parameters extracted in the same event multiplicity bins from a $\pi\pi$ femtoscopic study by 
ALICE \cite{Aamodt:2011kd} in 7 TeV $pp$ collisions. The K$^0_s$K$^0_s$ results 
for $R_{inv}$ suggest a slight tendency for $R_{inv}$ to decrease with increasing $k_T$
(or $m_T$) and to increase for increasing event multiplicity bin as also seen in 
the ALICE $\pi\pi$ results for 7 TeV $pp$ collisions and in heavy-ion collisions \cite{Adams:2004yc}.
Comparing with $\pi\pi$, the K$^0_s$K$^0_s$ results for $R_{inv}$ 
extend the covered range of $k_T$ to $\sim2$ GeV/c, which is about three times larger than
for $\pi\pi$. It is also seen in Figure \ref{fig5} that there is no discontinuity for the $k_T$ and
$m_T$ dependences of $R_{inv}$ between $\pi\pi$ and K$^0_s$K$^0_s$.

The author wishes to acknowledge financial support from the U.S.
National Science Foundation under grant PHY-0970048,  and to acknowledge computing
support from the Ohio Supercomputing Center.

\end{document}